# Project Icarus: Astronomical Considerations Relating to the Choice of Target Star

I. A. Crawford

Department of Earth and Planetary Sciences, Birkbeck College London, Malet Street, London, WC1E 7HX

## Abstract

In this paper we outline the considerations required in order to select a target star system for the Icarus interstellar mission. It is considered that the maximum likely range for the Icarus vehicle will be 15 light-years, and a list is provided of all known stars within this distance range. As the scientific objectives of Icarus are weighted towards planetary science and astrobiology, a final choice of target star(s) cannot be made until we have a clearer understanding of the prevalence of planetary systems within 15 light-years of the Sun, and we summarize what is currently known regarding planetary systems within this volume. We stress that by the time an interstellar mission such as Icarus is actually undertaken, astronomical observations from the solar system will have provided this information. Finally, given the high proportion of multiple star systems within 15 light-years (including the closest of all stars to the Sun in the α Centauri system), we stress that a flexible mission architecture, able to visit stars and accompanying planets within multiple systems, is desirable. This paper is a submission of the **Project Icarus Study Group**.

### 1. Introduction

The Icarus study is tasked with designing an interstellar space vehicle capable of making *in situ* scientific investigations of a nearby star and accompanying planetary system [1,2]. This paper outlines the considerations which will feed into the choice of the target star, the choice of which will be constrained by a number of factors. Foremost among these are the design criteria specified by the Project's Terms of Reference (ToR) document [1]. The most relevant ToR relating to the selection of the target star(s) are:

> ToR#3: "The spacecraft must reach its stellar destination within as fast a time as possible, not exceeding a century and ideally much sooner"; and
>
> ToR#5: "The spacecraft propulsion must be mainly fusion based"

Taken together, these imply a maximum realistic range of about 15 light-years from the Solar System. This would imply an interstellar cruise velocity of 15% of the speed of light (i.e. 0.15c), which is probably close to the upper end of what is feasible with a fusion-based propulsion system extrapolated from current knowledge. Moreover, given that ToR#3 adds the mission should 'ideally' be completed in less than 100 years, it follows that the ideal target would actually be significantly closer than 15 light-years.

For the purposes of this study we will therefore assume a maximum range of 15 light-years. ToR #4 states that “the spacecraft must be designed to allow for a variety of target stars”, from which it follows that the mission cannot be predicated on one particular target. Thus all stars within the maximum range need to be assessed as possible targets, although the wording of ToR#3 implies that, scientific considerations being equal, closer stars would be preferred.

## 2. Stars within 15 light-years of the Sun

Within 15 light-years of the Sun there are approximately 56 stars, in 38 separate stellar systems. The number is approximate for several reasons. Firstly, at the outer boundary the errors on the distances can amount to a few tenths of a light-year, which could mean that some stars notionally just beyond 15 light-years might actually be closer (and vice versa). Secondly, not all stars within this volume may yet have been discovered, although this is only likely for the very dimmest red or brown dwarfs. Thirdly, perhaps surprisingly, there are still slight discrepancies between the catalogues of nearby stars. Probably the most authoritative recent compilation, and the one on which the number of 56 stars is based, is the RECONS (Research Consortium on Nearby Stars) list of the one hundred nearest star systems [3].

The known stars within 15 light-years are listed in Table 1. Of these 56 stars, there is one star of spectral type A (Sirius); one F star (Procyon); 2 G stars (α Centauri A and τ Ceti); five K stars; 41 M stars (red dwarfs); 3 white dwarfs; and three probable brown dwarfs (the latter all members of multiple systems – there are no currently known free-floating brown dwarfs within this volume; these would be difficult to detect but in principle could exist).

**Table 1.** List of star systems within 15 light-years of the Sun [3]. GJ is each star’s number in the Gliese-Jahreiß catalogue [4]; *l,b* are each star’s galactic longitude and latitude, respectively.

| Dist. Order | GJ | Popular Name | Spectral Type | Distance (lt-yrs) | *l* (deg) | *b* (deg) |
|---|---|---|---|---|---|---|
| 1 | 551 | Proxima Cen | M5.5V | 4.2 | 313.9 | -01.9 |
| | 559 | α Cen A<br>α Cen B | G2V<br>K0V | 4.4 | 315.7 | -00.7 |
| 2 | 699 | Barnard’s Star | M4V | 6.0 | 031.0 | +14.1 |
| 3 | 406 | Wolf 359 | M6V | 7.8 | 244.1 | +56.1 |
| 4 | 411 | Lalande 21185 | M2V | 8.3 | 185.1 | +65.4 |
| 5 | 244 | α CMa A (Sirius)<br>α CMa B | A1V<br>DA2 | 8.6 | 227.2 | -08.9 |
| 6 | 65 | Luyten 726-8 A<br>Luyten 726-8 B | M5.5V<br>M6V | 8.7 | 175.5 | -75.7 |
| 7 | 729 | Ross 154 | M3.5V | 9.7 | 011.3 | -10.3 |
| 8 | 905 | Ross 248 | M5.5V | 10.3 | 110.0 | -16.9 |
| 9 | 144 | ε Eri | K2V | 10.5 | 195.8 | -48.1 |
| 10 | 887 | Lacaille 9352 | M1.5V | 10.7 | 005.1 | -66.0 |
| 11 | 447 | Ross 128 | M4V | 10.9 | 270.1 | +59.6 |
| 12 | 866 | EZ Aqr A<br>EZ Aqr B<br>EZ Aqr C | M5V<br>M?<br>M? | 11.3 | 047.1 | -57.0 |
| 13 | 280 | α CMi A(Procyon)<br>α CMi B | F5IV-V<br>DA | 11.4 | 213.7 | +13.0 |
| 14 | 820 | 61 Cyg A<br>61 Cyg B | K5V<br>K7V | 11.4 | 082.3 | -05.8 |
| 15 | 725 | Struve 2398 A<br>Struve 2398 B | M3V<br>M3.5V | 11.5 | 089.3 | +24.2 |
| 16 | 15 | (A) GX And<br>(B) GQ And | M1.5V<br>M3.5V | 11.6 | 116.7 | -18.4 |
| 17 | 845 | ε Ind A<br>ε Ind B<br>ε Ind C | K5Ve<br>T1<br>T6 | 11.8 | 336.2 | -48.0 |
| 18 | 1111 | DX Can | M6.5V | 11.8 | 197.0 | +32.4 |
| 19 | 71 | τ Cet | G8V | 11.9 | 173.1 | -73.4 |
| 20 | 1061 | | M5.5V | 12.0 | 251.9 | -52.9 |
| 21 | 54.1 | YZ Cet | M4.5V | 12.1 | 149.7 | -78.8 |
| 22 | 273 | Luyten’s Star | M3.5V | 12.4 | 212.3 | +10.4 |
| 23 | | Teegarden’s Star | M7V | 12.5 | 160.3 | -37.0 |
| 24 | | SCR1845-6357 A | M8.5V | 12.6 | 331.5 | -23.5 |
| | | SCR1845-6357 B | T | | | |
| 25 | 191 | Kapteyn's Star | M1.5V | 12.8 | 250.5 | -36.0 |
| 26 | 825 | AX Mic | M0V | 12.9 | 003.9 | -44.3 |
| 27 | 860 | Kruger 60 A<br>Kruger 60 B | M3V<br>M4V | 13.1 | 104.7 | +00.0 |
| 28 | | DEN J1048-3956 | M8.5V | 13.2 | 278.7 | +17.1 |
| 29 | 234 | Ross 614 A<br>Ross 614 B | M4.5V<br>M8V | 13.3 | 212.9 | -06.2 |
| 30 | 628 | Wolf 1061 | M3.0V | 13.8 | 003.4 | +23.7 |
| 31 | 35 | Van Maanen's Star | DZ7 | 14.1 | 121.9 | -57.5 |
| 32 | 1 | | M3V | 14.2 | 343.6 | -75.9 |
| 33 | 473 | Wolf 424 A<br>Wolf 424 B | M5.5V<br>M7V | 14.3 | 288.8 | +71.4 |
| 34 | 83.1 | TZ Ari | M4.5 | 14.5 | 147.7 | -46.5 |
| 35 | 687 | | M3V | 14.8 | 098.6 | +32.0 |
| 36 | 3622 | LHS 292 | M6.5V | 14.8 | 261.0 | +41.3 |
| 37 | 674 | | M3V | 14.8 | 343.0 | -06.8 |
| 38 | 1245 | V1581 Cyg A<br>V1581 Cyg B<br>V1581 Cyg C | M5.5V<br>M6.0V<br>M? | 14.8 | 078.9 | +08.5 |

## 3. Scientific criteria for the choice of target star

The scientific objectives of an interstellar vehicle such as Icarus have been described by Webb [5] and Crawford [6], and for Icarus specifically as a trade study conducted within Icarus Module 14 [7]. Briefly, these scientific objectives can be divided into the following broad categories:

(1) Science to be conducted on route, e.g. of the local interstellar medium (LISM), and physical and astrophysical studies which could make use of the Icarus vehicle as an observing platform;

(2) Astrophysical studies of the target star itself, or stars if a multiple system is selected;

(3) Planetary science studies of any planets in the target system, including moons and large asteroids of interest;

(4) Astrobiological/exobiological studies of any habitable (or inhabited) planets or moons which may be found in the target planetary system.

In order to help identify criteria which may be used to help select the choice of target stars(s) for Icarus, we propose here that the above areas of scientific investigation be considered as listed in order of increasing priority. Thus, scientific investigations conducted en route are a low priority when it comes to the choice of target, not because such investigations are scientifically unimportant but because they can (largely) be conducted regardless of what the target star is chosen to be. It is true that in some directions the local interstellar medium is of more interest than others [6, 8], but this is unlikely to be a scientific driver for a vehicle as complex and costly as an Icarus-type starship.

Astrophysical studies of the target star(s) will have a higher priority. Although all potential targets will be nearby stars, about which much can be learned from astronomical observations from the solar system, detailed studies of, e.g., photospheric structure, magnetic properties, and stellar wind, would clearly benefit from the possibility of *in situ* observations [6]. From this perspective, higher priority might be given to rare or unusual stars (as argued by Martin [9] in the context of the Daedalus study). Examples might include an early type star and/or a white dwarf (both might be achieved by selecting either Sirius or Procyon as a target; Table 1), a brown dwarf (which could be achieved by visiting the ε Indi system), or a nearby M dwarf (which, although not rare, are perhaps the least understood class of main sequence stars, in part owing to their intrinsic faintness). That said, one should not underestimate the scientific importance of making *in situ* observations of another main-sequence G-type star (such as α Centauri A or τ Ceti) to allow comparisons with the Sun. However, interesting and important as these astrophysical considerations are, by themselves they also are unlikely to be the main scientific drivers for an interstellar space mission such as Icarus.

On the other hand, there seems little doubt that the presence of a planetary system will greatly increase the scientific priority of a potential target star. This is because there are many aspects of planetary science which can only be addressed by *in situ* measurements, including the landing of scientific instruments on planetary surfaces [6]. In addition to the intrinsic planetary science interest in such objects, the habitability of any such planets will be of compelling scientific interest. This will be especially true if astronomical observations from the Solar System indicate that any may actually be inhabited. In the latter case, definitive proof of the existence of indigenous life, and studies of its underlying biochemistry, cellular structure, ecological diversity and evolutionary history will require *in situ* measurements to be made [6]. Thus it seems clear that, when it comes to selecting a final target star system for Icarus, the presence of a planetary system, and especially the presence of habitable or inhabited planets, will trump all other scientific motivations.

Given the importance of planets in selecting an astronomical target for Icarus, in the next Section we review the evidence for planets around the nearest stars.

## 4. Planets within 15 light-years of the Sun

An excellent summary of all known extrasolar planets (currently more than 500) can be found in the Extrasolar Planet Encyclopedia maintained by Jean Schneider [10]. Two of the stars in Table 1 are already known to have planets, on the basis of radial velocity measurements. These are ε Eridani (a single K2 star at a distance of 10.5 light-years), and the M3 red dwarf GJ 674 at a distance of 14.8 light-years. In addition, there are strict observational upper-limits on the masses of any planets which could exist orbiting three other M dwarfs listed in Table 1, namely Proxima Centauri, Barnard's Star, and GJ1. These specific cases are briefly discussed in this Section. For completeness, we note that there are a couple of other M dwarfs, just beyond the 15 light-year range considered feasible for Icarus, which are also known to have planets, namely GJ 876 (15.3 light-years) and GJ 832 (16.1 light-years), but these will not be considered further here.

### 4.1 ε Eridani

The planet orbiting ε Eri is a giant planet, with a mass about 1.5 times that of Jupiter (i.e. 1.5 $M_J$; [11]). It has a highly eccentric orbit, which brings it as close to its star as 1.0 AU (i.e. the same distance as the Earth is from the Sun), to as distant as 5.8 AU (i.e. just beyond the orbit of Jupiter in our Solar System), with a period of 6.8 years. Although this would span the habitable zone (i.e. the range of distances from a star on which liquid water would be stable on a planetary surface given certain assumptions about atmospheric composition), this orbit lies wholly outside the likely habitable zone for a K2 star like ε Eri. Also, being a gas giant, this planet itself it not a likely candidate for life, and its eccentric orbit wouldn't help in this respect (although it is possible that the planet may have astrobiologically interesting moon's, perhaps similar to Jupiter's moon Europa, which could in principle support sub-surface life).

There is an unconfirmed detection of another planet in the ε Eri system, of intermediate mass (0.1 $M_J$) in a very distant (40 AU) orbit [12]. It is possible that the system contains lower mass, more Earth-like, planets, which might be more interesting targets for investigation, especially closer to the star than the giant planet that is known to exist. ε Eri is also known to be surrounded by a disk of dust [13], which may be derived from collisions between small planetesimals (i.e. asteroids and/or comets), which is an indirect argument for smaller planets also being present. Only further research will tell how many planets actually reside in the ε Eri system, and whether any are of astrobiological interest. The existence of at least one planet, and the dust disk (itself of great astrophysical interest), would make epsilon Eri a high priority candidate target for Icarus if it were not for its distance of 10.5 light-years. Although within the 15 light-year radius considered here, this is still a very challenging distance for the first attempt at an interstellar voyage.

### 4.2 GJ 674

At a distance of 14.8 light-years GJ 674 is right on the limit of the distance range considered here. The planet orbiting this star is very different from those orbiting ε Eri -- with a mass of only about 11 Earth masses (i.e. 11 $M_E$; [14]) it is likely to be a giant rocky planet: a so-called 'super-Earth'. It orbits its star every 4.7 days, in a moderately elliptical orbit at a mean distance of only 0.04 AU (one tenth of Mercury's distance from the Sun!). Even for a red dwarf star, this is probably too close to be habitable. However, as one planet exists around this star it is possible that others will be discovered, perhaps in more habitable orbits, as observations continue. Only time will tell, but in any case the distance of this star probably renders it of marginal interest for Icarus.

### 4.3 Other observational limits on planets within 15 light-years

A recent study by Zechmeister et al. [15] reports non-detections of planets for three stars listed in Table 1. In order of increasing distance these are: Proxima Centauri, Barnard's Star, and GJ 1. The corresponding upper mass (strictly m×sin(*i*), where *i* is the unknown orbital inclination) limits for planets orbiting within 2.5 AU of these stars are about 0.1, 0.1, and 0.3 $M_J$, respectively (i.e. planets more massive than these limits would have been detected; see Figure 6 of Reference [15]). The upper mass limits for planets orbiting closer to these stars (i.e. within the appropriate M dwarf 'habitable zone') are more restrictive: approximately 2.5-4, 3-6, and 20-30 Earth-masses for Proxima Cen, Barnard's Star and GJ1, respectively. Note that although these observations rule out massive planets within a few AU, consistent with the general rarity of such planets around M dwarfs (discussed further in Section 5.1), and 'super-Earth-like' planets in shorter period orbits, they do not yet exclude Earth- or sub-Earth-mass planets in the habitable zone (which would be below the sensitivity of the current measurements). The statistical results discussed in Section 5 actually imply that it is quite likely that one or more of these nearby M dwarf stars will be found to be accompanied by one or more low-mass planets. Only further observations will tell.

Other stars in Table 1 are also subject to careful scrutiny by the various radial velocity surveys, especially the solar type stars α Cen A/B [16] and τ Cet. Although at the time of writing specific planetary mass limits have not yet been published, it is only a matter of time before firm limits on the presence of at least Jupiter-mass planets are available for these stars. The presence of a significant dust disk orbiting τ Cet [17] does seem to imply the presence of some kind of planetary (or at least cometary) system, even if giant planets are absent.

## 5. Statistical properties of exoplanets: implications for the probable number of planets within 15 light-years

Clearly it would be of great interest if planets were discovered orbiting stars closer than ε Eri and GJ 674. To-date no such planets have been discovered, but they are very likely to exist. The statistics on the prevalence of planetary systems are not yet complete, but it is already becoming clear that several tens of percent of stars are accompanied by a planetary system of one kind or another. That said, it is also becoming clear that the distributions of low and high-mass planets, and their association with stars of different spectral types, are probably rather different. In this section we attempt to summarise the results of recent planet surveys which may inform estimates of the prevalence of planets within 15 light-years of the Solar System.

### 5.1 High mass (0.3-10 $M_J$) planets

The situation with regard to massive planets (i.e. 0.3 to 10 Jupiter masses) orbiting broadly solar type stars (spectral classes FGK) with orbital periods in the range 2 to 2000 days (corresponding to semi-major axes of 0.03 and 3 AU for a solar mass star) has been summarised by Cumming et al. [18]. The statistics are reasonably complete for this range of mass and orbital period, and Cumming et al. find that 10.5% of solar type stars are accompanied by such giant planets. For longer orbital periods, which would include true Jupiter analogues, the data are less complete, although Wittenmyer et al [19] have recently reported on a search for these, and find that at least 3.3% of solar type stars are accompanied by giant planets with 'Jupiter-like' (i.e. 3 to 6 AU) orbits. This latter result is consistent with an extrapolation of the results of Cumming et al. [18] who predict that 2.7±0.8 percent of solar type stars will have a giant planet in this distance range. Combining these two results yields an estimate that about 14% of solar type stars have a giant planet with orbital semi-major axes up to 6 AU. A (possibly unwarranted) extrapolation of the results of Cumming et al. [18] predicts that between 17 and 20 percent of solar-type stars will have a giant (0.3 to 10 $M_J$) planet orbiting within 20 AU; it will take several more decades of observations to determine whether or not this latter prediction is correct owing to the long orbital periods of such planets.

We note that, as far as Icarus is concerned, the statistical prevalence of planets orbiting solar-type stars is of marginal relevance, as only 8 out of the 56 stars listed in Table 1 (14%) are of spectral types F, G or K. Indeed, based on the above statistics, we might only expect one of these eight stars to have a massive planet within 6 AU, which is (statistically) consistent with the known ε Eri planet(s) and the continuing lack of detections of such planets orbiting α Cen A/B or τ Ceti.

Given the overwhelming preponderance of M dwarfs within 15 light-years of the Sun (Table 1), it is the statistical prevalence of planets around this class of star which is of most interest for the Icarus project. Massive planets are easier to detect orbiting M dwarfs than for solar-type stars, principally because the low stellar mass results in a larger Doppler 'wobble' for a given planet mass. However, in site of this advantage, and the fact that several M dwarfs have been discovered to have planets (including GJ 674), it is becoming clear that giant planets in short period orbits are actually rarer around M dwarfs than they are around solar-type stars [15, 18]. This is at least qualitatively understandable because lower mass stars presumably form from lower mass protostellar/protoplanetary nebulae, so there is likely to be less mass available in the circumstellar environment from which giant planets might form. Specifically, Cumming et al. [18] estimate that only 2% of M dwarfs are accompanied by giant planet with orbital periods of under 2000 days (2.5 AU for a 0.5 solar-mass M dwarf), compared to their value of 10.5% for solar-type stars. This is consistent with the fact that a radial velocity survey of 40 M dwarf stars by Zechmeister et al. [15] failed to detect any planets, and would be consistent with there being no giant planets orbiting the 41 M dwarfs listed in Table 1.

However, there is an important caveat to add to this conclusion. Statistical studies of planets detected by gravitational microlensing, which is sensitive to massive planets in more distant orbits, has found these to be more common than the shorter period planets found by radial velocity surveys [20, 21]. Specifically, Gould et al. [20] find that the frequency of giant planets (roughly in the range 0.02 to 5 $M_J$) orbiting low-mass (~0.5 solar-mass) stars with orbital radii beyond the 'snow line' (~1.4 AU for a M0 star) is about a factor of 8 higher than found by the radial velocity surveys for shorter periods. Although there is an overlap in the range of orbital radii covered by these techniques, the microlensing results are biased towards the larger orbital distances, with greatest sensitivity at about three times the local 'snow line' (i.e. ~4 AU for a 0.5 solar-mass star given the assumptions of Gould et al. [20]). Gould et al. note that this increase in the frequency of planets with increasing orbital radii is consistent with the trend observed at shorter periods by Cumming et al [18] (see Figure 8 of Reference [20]), and predicts that the radial velocity surveys will start to detect these longer period planets orbiting low-mass stars once sufficient observational time has accrued. If the statistical results of Gould et al. [20] apply to stars in the solar neighbourhood, they would raise the estimate of the fraction of local M dwarf stars accompanied by massive planets from ~2% to

~18% (including the 2% already implied by the radial velocity results). Note that, from an astrobiological perspective, all these extra planets would be well beyond the local habitable zones for M dwarf stars, and therefore likely to be of planetary science interest only (although the possibility of life below the surfaces of attendant moons, such as that proposed for Jupiter's moon Europa [22], would need to be considered).

### 5.2 Lower mass (<0.1 $M_J$, or 30 $M_E$) planets

There is growing observational evidence that, despite being harder to detect, lower mass planets in short period orbits are more common around solar type stars than the high mass planets discussed in Section 5.1. Indeed, all radial velocity surveys reported to-date have found a strongly increasing planet occurrence with decreasing planet mass [23, 24, 25]. Mayor et al. [23] estimate that 30% of solar-type (i.e. FGK) stars are accompanied by low mass (<30 $M_E$) planets with periods <100 days (i.e. semi-major axes ≤0.4 AU). This is consistent with the recent results of Wittenmyer et al. [24], who find 18.5% of solar-type stars with planets <10$M_E$ with periods <50 days, and Howard et al. [25], who find 18.3% of solar-type stars with planets in the range 3-30 $M_E$ also with periods <50 days. Extrapolation of the results of Howard et al. to lower masses implies 28% of solar-type stars have planets in the mass range 0.5 to 3 $M_E$ (with 23% in the range 0.5 to 2 $M_E$) bringing their estimated total occurrence of planets in the mass range 0.5 to 30 $M_E$ with periods < 50 days to 46.3%. Extending these results to low-mass planets in longer period (i.e. more Earth-like) orbits is a high priority for future searches, but pushes the sensitivity of current techniques owing to the low masses of these planets and the decrease in radial velocity amplitude with increasing orbital distance.

Although not targeted at any of the nearest stars, results from the Kepler mission [26], which is looking for low-mass planets orbiting solar-type stars by the transit method, will greatly improve these estimates within the next few years. Already, after correction for known sensitivity biases, the preliminary results [27] indicate that planetary candidates with orbital periods <138 days (i.e. orbital semi-major axes <0.5 AU for a solar-mass star) are common. Within this range of orbital periods, the preliminary Kepler estimates for the fraction of stars with planets in different size ranges are as follows: Earth-size planets (<1.25 $R_E$, where $R_E$ is the Earth's radius): 5.4%; Super-Earth-size planets (1.25 – 2.0 $R_E$): 6.8%; Neptune-size planets (2.0 – 6.0 $R_E$): 19.3%; Jupiter-size planets (6.0 – 15 $R_E$): 2.4%; and, finally, very large planets (>15 $R_E$): 0.15%. This gives a total estimated abundance of planets with orbital periods <138 days of 34%. Considering the uncertainties involved, and considering the aggregate number of planets across all size ranges, the preliminary Kepler results are in reasonable agreement with the results based on the radial velocity measurements, with about 30±10% of stars being accompanied with planets in relatively short period orbits. As for the radial velocity results, and for essentially the same reasons, the corresponding statistics for longer period, more Earth-like, orbits will take several years to accumulate.

That said, there are a couple of caveats. Firstly, to-date, Kepler has found only about half the fraction of planetary candidates in the 'Earth' to 'Super-Earth' (i.e. 1 to 8 $M_E$, or 1-2 $R_E$) range (12.2%) than predicted by an extrapolation of Howard et al.'s [25] planetary mass distribution (23%), and it may be that the latter overestimates the occurrence of such planets in short (< 50 day) orbits. Secondly, and most importantly, the Kepler observations include stars of all spectral types and are not restricted to the FGK spectral classes which dominate most of the radial velocity surveys. Kepler data therefore enables us to say something about the prevalence of low-mass planets around M dwarf stars, which dominate the immediate solar neighbourhood (Table 1). Reassuringly, these results confirm the relative paucity of Jupiter-mass planets around M dwarfs found by the radial velocity surveys, but indicate that lower mass (i.e. 'Neptune' and 'Super-Earth' sized) planets are actually *more* common around M dwarfs than around FGK stars (see Figure 15 of Reference [27]), while the abundance of the smaller 'Earth'-sized planets is about the same around M dwarfs as for G stars (with K stars showing a possibly significant deficit). Considering all planetary size ranges, the preliminary Kepler results indicate that planets are more common around M dwarfs than around any other class of star, with the following estimated abundances for planets with orbital periods <138 days as a function of stellar spectral type: F: 33%; G: 26%; K: 24%; and M: 48% [27]. However, it is important to stress the preliminary nature of these results – further observations are required to confirm these tentative conclusions regarding the occurrence of low-mass planets as a function of stellar type.

### 5.3 Implications of statistical results for the prevalence of planets within 15 light-years of the Sun

Bearing in mind the observational uncertainties, and the provisional nature of the Kepler results, it seems safe to conclude that at least 30±10% of the stars in Table 1 will be accompanied by planetary systems. While only a small fraction (~2%) of the local M dwarfs are likely to be accompanied by a giant (Jupiter-mass) planet within several AU, this is probably compensated by a higher fraction (perhaps as large as ~50%) of these stars being accompanied by lower mass planets in short (<138 day) orbits. Both estimates are strict lower limits owing to incompleteness for longer orbital periods, especially for the low-mass planets where the results to-date do not include planets in Earth-like (~1 AU) or more distant orbits. On the basis of statistics collected to-date, it therefore seems safe to conclude that at least 17±6 of the stars listed in Table 1 will be accompanied by a planetary system of one sort or another.

The actual number is likely to be considerably higher than this, owing to incompleteness in both the ground-based radial velocity and the Kepler transit datasets. In particular, the statistical results form gravitational microlensing [20, 21] imply that massive planets are relatively common with orbital periods longer than those yet probed by the radial velocity surveys, and the provisional Kepler results for low-mass planets [27] do not yet include those with orbital periods longer than about 140 days (i.e. orbital radii >0.5 AU) yet it would be extraordinary if such planets are not found as the Kepler mission proceeds.

## 6. Characterising nearby exoplanets as potential targets for Icarus

Although the statistical arguments described above demonstrate the likelihood that planets will eventually be found orbiting a large fraction of stars within 15 light-years of the Sun, when the time comes to finally select a specific target star for Icarus it will be necessary for this system to be well characterized in advance. Indeed, such characterization will likely be required in order to decide between competing target star systems. Fortunately, as already noted by Long et al. [2] and Crawford [6], well before rapid interstellar space travel becomes possible, advances in astronomical techniques will have already identified which of the nearest stars are accompanied by planetary systems. Indeed, we are likely to know the basic architecture of these systems in some detail, and solar system-based instruments will have the capability of detecting any molecular biosignatures that may be present in the atmospheres and/or on the surfaces of these planets [28, 29, 30].

We can therefore be confident that astronomical observations will be able to establish a hierarchy of priorities among any planets which may be detected around the nearest stars: (i) planets where *bona fide* biosignatures are detected; (ii) planets that appear habitable (e.g. for which there is spectral evidence for water and carbon dioxide, but no explicit evidence of life being present); and (iii) planets which appear to have uninhabitable surfaces (either because of atmospheric compositions deemed non-conducive to life or because they lack a detectable atmosphere), but which might nevertheless support a subsurface biosphere. Thus, when planning an interstellar mission with astrobiology/exobiology in mind, we are likely to have a priority list of target systems prepared well in advance.

## 7. The special case of the α Centauri system

The relative proximity of α Cen, together with its especially interesting interstellar sightline [6, 8] and the presence of stars of three different spectral types, makes it an attractive target for humanity's first interstellar mission. However, as the bulk of the scientific benefits of interstellar spaceflight pertain to planetary science and astrobiology (Section 3), a final prioritization must await future developments in the detection of planets in the system. This should be forthcoming within the next few years.

It is interesting to note that, although the Daedalus study was predicated on a mission to Barnard's Star, the ranking of stellar targets conducted for that study by Martin [9] actually found α Cen to be by far the highest priority target among the nearest stars. Although this conclusion was reached on the basis of a numerical weighting scheme which seems difficult to justify given the lack of information regarding planets around these stars, it is difficult to see how any rational application of the criteria sketched in Section 3 could come to a different conclusion. The fact is that of all the nearest stars the α Cen system offers the most diverse path through the local interstellar medium [6, 8], and the greatest diversity of stellar spectral types. Only if the α Cen system is found not to contain any planets, whereas another nearby star is found to harbour a habitable, or even inhabited, planetary system is α Cen likely to lose its place at the top of the priority list of nearby stellar targets.

Although the Icarus ToR [1, 2] prevent the mission being designed specifically for a single target star, the high priority likely to be attached to the α Cen system means that some consideration should be given to implementing a mission architecture capable of exploring double or multiple star systems. In the particular case of α Cen, if planets are discovered around either (or both) components A and B the ideal architecture would be one which decelerates into the A/B system, but which also launches an undecelerated flyby probe to Proxima Cen (located 2.18 degrees away on the sky). On the other hand, should Proxima Cen be discovered to harbour a planetary system, and α Cen A/B not, then it may be appropriate to decelerate at the Proxima system and send a flyby probe to α Cen A/B. The practicalities of such a mission architecture should be considered in future studies [6, 7]. Given the high proportion of binary and multiple star systems listed in Table 1, a flexible mission architecture of this kind is likely to be useful even if α Cen is not chosen as the target star system.

## 8. Conclusions

Only further observational work will reveal how common planets actually are around the nearest stars. The expectation based on statistical results from radial velocity surveys, and provisional results from the Kepler transit survey, is that several tens of percent (roughly 30±10%) of stars will be accompanied by a planetary system of some kind. This is likely to be a lower limit, as the statistics relating to key areas of the planetary mass – orbital distance parameter space are not yet complete (and are unlikely to be complete for several decades). On this basis we might expect (at least) 17±6 of the stars listed in Table 1 to have planets. For the M dwarf population which dominates the solar neighbourhood (Table 1) giant planets are likely to be quite rare, but low mass (i.e. 'Earth-', and 'super-Earth'-mass) planets may be common.

Before a final choice of target can be selected for an Icarus-type interstellar mission, it will be necessary to fully characterize the existence and nature of planetary systems in the immediate solar neighbourhood. Fortunately, long before we are able to build an Icarus-type starship, astronomical technology will have reached the point where we are likely to have a complete census of planetary systems within 15 light-years of the Sun. Not only will these instruments be able to identify which stars have planets, and calculate their orbital parameters, they will be able to make basic spectroscopic searches for biosignatures in their atmospheres or on their surfaces. Thus, although currently we cannot identify an obvious specific target for Icarus (other than draw attention to the currently perceived advantages of the α Cen system), when the time comes to actually build a starship such as Icarus we will have a very good idea where to send it. Given the high proportion of binary and multiple star systems within 15 light-years of the Sun, a flexible mission architecture able to visit two or more stars, and their accompanying planets, within such multiple systems would be desirable.